\def\kms{km~s$^{-1}$}
\def\sm{$M_\odot$}
\documentclass{aa501}
\usepackage{graphicx}

\begin{document}

\title{Evidence for binarity in the bipolar planetary nebulae \object{A~79},
\object{He~2-428} and \object{M~1-91}
\thanks{Based on observations made with the
2.5~m INT and the 2.6~m NOT telescopes operated on the island of La
Palma by the ING and the NOTSA, respectively, in the Spanish
Observatorio del Roque de Los Muchachos (ORM) of the Instituto de
Astrof\'\i sica de Canarias.}}

\author{M. Rodr\'\i guez\inst{1}, R. L. M. Corradi\inst{2} and
	A. Mampaso\inst{3}}

\offprints{M.~Rodr\'\i guez}

\institute{Instituto Nacional de Astrof\'\i sica, \'Optica
y Electr\'onica INAOE, Apdo Postal 51 y 216, 72000 Puebla, Pue., M\'exico
(mrodri@inaoep.mx)
\and
Isaac Newton Group of Telescopes, Apartado de Correos 321, 38700 Santa 
Cruz de La Palma, Canarias, Spain (rcorradi@ing.iac.es)
\and
Instituto de Astrof\'\i sica de Canarias, E-38200 La Laguna, 
Tenerife, Canarias, Spain (amr@ll.iac.es)}

\date{Received date; accepted date}

\abstract{
We present low and high resolution long-slit spectra of three bipolar
planetary nebulae (PNe) with bright central cores: \object{A~79},
\object{He~2-428} and \object{M~1-91}.
\object{He~2-428} and \object{M~1-91} have high density (from $10^{3.3}$ to
$10^{6.5}$~cm$^{-3}$) unresolved nebular cores that indicate that
strong mass loss/exchange phenomena are occurring close to their
central stars.
An F0 star is found at the centre of symmetry of \object{A~79};
its reddening and distance are consistent with the association of the star
with the nebula.
The spectrum of the core of \object{He~2-428} shows indications of the presence
of a hot star with red excess emission, probably arising in a late-type
companion.
\object{A~79} is one of the richest PNe in \element{N} and \element{He}, the
abundances of \object{M~1-91} are at the lower end of the range
spanned by bipolar PNe, and \object{He~2-428} shows very low
abundances, similar to those measured for halo PNe.
The extended nebulae of \object{A~79} and \object{He~2-428} have
inclined equatorial rings expanding at a velocity of $\sim$15~\kms,
with kinematical ages $\ge$10$^4$~yrs.  The association of these aged,
extended nebulae with a dense nebular core (He~2-428) or
a relatively late type star (A~79) is interpreted as evidence for the
binarity of their nuclei.
}

\titlerunning{\object{A~79}, \object{He~2-428} and \object{M~1-91}}
\authorrunning{M.~Rodr\'\i guez et al.}

\maketitle

\section{Introduction}

Most planetary nebulae (PNe) possess axisymmetrical shapes, and some
10--15\% of the global sample display a marked bipolar morphology, in
the form of an `equatorial' waist or ring from which symmetrical lobes
depart (Corradi \& Schwarz \cite{CS95}).

It has been shown that pure hydrodynamical collimation provided by
dense equatorial disks or torii (Icke et al.\ \cite{I89}), and/or
magneto-hydrodynamical collimation requiring the presence of toroidal
fields (Chevalier \& Luo \cite{che94}) can explain the development of the
extreme bipolar geometries observed.  While the dynamical mechanisms
leading to the formation of bipolar lobes have therefore been
identified, the AGB or post-AGB phase at which the asymmetries develop
(e.g. Sahai \cite{sah00}) and the ultimate cause of their occurrence are
still not well understood.  In particular, whether a binary
companion is needed to cause or enhance the asymmetry in the wind of
the PN progenitor is the subject of a long term debate (cf. the
proceedings of the recent conference edited by Kastner, Soker \&
Rappaport \cite{kas00}). A recent view (Soker \& Rappaport \cite{sok00})
is that bipolar PNe, especially those with the most extreme morphologies, are
formed in binary systems that avoid a common envelope evolution during
the AGB phase of the star on the way of producing the PN, i.e. by
binaries with (present) orbital separations from $\sim$5 to
$\sim$200~AU.
It is hard to find evidence of binary motions from radial
velocity measurements, especially for the longest period systems.
An additional, major reason that makes it
difficult to find a direct evidence for binarity in bipolar PNe is
that their central stars are generally faint.  In some cases, this is
because of dust obscuration in dense circumstellar discs surrounding
the central stars, but in general it is due to the fact that bipolar PNe
are produced by the most massive progenitors of PNe ($M\ge2.3$~\sm,
Phillips \cite{phi01}), which rapidly evolve towards
low-luminosities in their post-AGB evolution.

For these reasons, the surprisingly high brightness of the central star
of \object{He~2-428} (PN G049.4+02.4), whose bipolar nebula appears
rather old in the existing narrowband images, attracted our attention
and we decided to investigate its nature by means of low and high
resolution spectroscopy.
\object{Abell~79} (PN G102.9$-$02.3) and \object{M~1-91} (PN G061.3+03.6)
were also observed because they share some similarities with
\object{He~2-428} and with a subclass of bipolar PNe having properties
reminding those of known binary systems such as symbiotic stars
(Corradi \cite{cor95}). The results of our spectroscopic study are
presented in this paper.

\section{Observations and data analysis}

\subsection{Low-resolution spectroscopy}

\begin{table*}
\caption[ ]{Journal of observations}
\begin{tabular}{lllllll}
\hline
\noalign{\smallskip}
Object & $\alpha(2000)$ & $\delta(2000)$ & PA & Date & Exposure times &
	Scale      \\
 & (hh~mm~ss) & (\degr~\arcmin~\arcsec) & (\degr) & &  (min.) &
 	(\arcsec~pixel$^{-1}$)\\
\noalign{\smallskip}
\hline
\noalign{\smallskip}
\multicolumn{7}{c}{Low-resolution spectroscopy} \\
\noalign{\smallskip}
\hline
\noalign{\smallskip}
\object{He~2-428} & 19~13~05 & 15~46~42 & 0 & 1996~July~25 &
	30, 3 ($\lambda\lambda$3690--6300) & 0.41 \\
           &   &   &   &   & 20, 3 ($\lambda\lambda$6020--9220) &  \\
\object{M~1-91} & 19~32~57 & 26~52~40 & $+72$ & 1996~July~25 &
	30, 3 ($\lambda\lambda$3690--6300) & 0.41 \\
           &   &   &   &   & 20, 3 ($\lambda\lambda$6020--9220) &  \\
\object{A~79} & 22~26~17 & 54~49~38 & $-20$ & 1996~Aug~3 &
	$30\times2$, $3\times2$ ($\lambda\lambda$3690--7050) & 0.84 \\
\noalign{\smallskip}
\hline
\noalign{\smallskip}
\multicolumn{7}{c}{High-resolution spectroscopy}  \\
\noalign{\smallskip}
\hline
\noalign{\smallskip}
\object{He~2-428} & & & $-10$ & 1999~August~31 & $30\times2$ & 0.14 \\
                  & & & $+80$ &                & $30$        &      \\
\object{A~79}     & & & $+33$ & 1999~August~31 & $30$        & 0.14 \\
                  & & & $-57$ &                & $30$        &      \\
\noalign{\smallskip}
\hline
\end{tabular}
\end{table*}

Long-slit spectra covering the effective ranges
$\lambda\lambda$3650--9220 (\object{He~2-428} and \object{M~1-91}) and
$\lambda\lambda$3690--7050 (\object{A~79}) were obtained at the 2.5~m
Isaac Newton Telescope on La Palma equipped with the IDS spectrograph,
the 500~mm camera (\object{He~2-428} and
\object{M~1-91}) or the 235~mm camera (\object{A~79}) and a CCD detector.
The spectral ranges were covered with spectral resolutions of $\sim8$~\AA\
using either a grating of 150~lines~mm$^{-1}$ at two angles (\object{He~2-428}
and \object{M~1-91}), or a grating of 300~lines~mm$^{-1}$ at one angle
(\object{A~79}) and slit widths of 1\farcs1 (\object{He~2-428} and
\object{M~1-91}) and 1\farcs6 (\object{A~79}).
The long slit was placed through the central bright cores of the objects and
across the rings and lobes.
The positions and orientations (PA) of the slit, the spatial resolutions and
the exposure times are listed in Table~1.

Bias frames, twilight or tungsten flat-field exposures, wavelength
calibrations and exposures of the standard star \object{BD~+284211}
(from the IRAF compilation) were taken each night. The spectra were
reduced to absolute intensity units using the IRAF reduction package,
following standard procedures for the long-slit case.
After the bias subtraction, flat-field correction and two-dimensional
wavelength calibration, the individual exposures were flux calibrated
using the standard star fluxes and the mean extinction curve for La
Palma.  Cosmic rays were removed and, after subtracting the sky
contribution, the equivalent nebular spectra were combined to improve
the signal-to-noise ratio.

One dimensional spectra were then extracted over the main
morphological regions of the nebulae: the core or central star, the
lobes and/or rings.  The two regions on each side of the central star are
identified in the following as A (the region towards the north for \object{A~79}
and \object{He~2-428}, and the region towards the west for \object{M~1-91})
and B (the regions towards the south or east).
Tables 2-6 give the positions and sizes of the selected regions.
Line intensities were measured in each region by fitting Gaussian profiles.

\subsection{High-resolution spectroscopy}

High resolution, long-slit spectra of \object{He~2-428} and
\object{A~79} were obtained 
at the 2.6~m Nordic Optical Telescope on La Palma, using the echelle
spectrograph IACUB and a Thompson THX31156 CCD.
The projected slit width was 0\farcs5, providing a spectral resolution
$R=\lambda/\Delta\lambda=$40000, with a reciprocal dispersion of
0.05~\AA\ per binned pixel.  The spectral range, selected using a
narrowband filter, includes H$\alpha$ and the
[\ion{N}{ii}]~$\lambda6583$ line.
The slit position angles and exposure times are given in Table~1.
The data were reduced using standard procedures in IRAF.
The uncertainties in the measured
relative velocities are of $\sim$2~km~s$^{-1}$.

\subsection{Imaging}

An $\element{H}\alpha+[\ion{N}{II}]$ image of He~2-428 was obtained on
April 25, 2001, at the prime
focus Wide Field Camera of the INT, equipped with a mosaic of 4 EEV
$2048\times4100$ CCDs.
The scale was 0\farcs33, the exposure time 10~min and the seeing
1\farcs1.  The central wavelength and width of the
$\element{H}\alpha+[\ion{N}{II}]$ filter are
6568~\AA\ and 95~\AA, respectively.

\subsection{Extinction correction and atomic data}
\label{atdata}

In the low resolution spectra, line intensities have been corrected
for extinction by interpolating Cardelli et al.'s (\cite{car89})
$R_{\rm V}$-dependent extinction law, with
$R_{\rm V}=A_{\rm V}/E_{\rm B-V}=3.1$, the mean
value for the diffuse interstellar medium.
The logarithmic extinction $c$ (see Osterbrock \cite{ost89}) has been
calculated for the outer nebular regions by fitting the ratio
$I(\mathrm{H}\alpha)/I(\mathrm{H}\beta)$ to its recombination value for
suitable physical conditions:
$I(\mathrm{H}\alpha)/I(\mathrm{H}\beta)=2.80\mbox{--}2.88$ for
$N_\mathrm{e}\simeq1000\,\mathrm{cm}^{-3}$ and
$T_\mathrm{e}=10\,000\mbox{--}15\,000\,\mathrm{K}$.
The procedures followed to determine the values of the extinction parameter
$c$ in the cores are explained in Sects~\ref{na79}, \ref{nhe2428} and
\ref{nm191}.
All the values used for $c$ are given in Tables 2--6, along with the measured
and reddening-corrected line intensities for each area.

The line ratios of [\ion{S}{ii}] lines $I(\lambda6716)/I(\lambda6731)$,
[\ion{N}{ii}] lines $I(\lambda6548+\lambda6583)/I(\lambda5754)$ and
[\ion{O}{iii}] lines $I(\lambda4959+\lambda5007)/I(\lambda4363)$ have been
used to determine densities and temperatures.
At high densities, like those found in the cores of \object{He~2-428} and
\object{M~1-91} (see below), the [\ion{N}{ii}] and [\ion{O}{iii}] ratios are
strongly dependent on density and can be used to determine this quantity if a
value for the temperature is assumed.
The physical conditions based on these lines as well as the ionic abundances
implied by all the collisionally excited lines, excluding [\ion{Fe}{iii}],
have been determined with the {\em nebular} package in IRAF, with the atomic
parameters referenced therein.
The \ion{He}{i} and \ion{He}{ii} emissivities have been taken from Benjamin et
al.\ (\cite{ben99}), [\ion{Fe}{ii}] emissivities are from Bautista \& Pradhan
(\cite{bau96}).
Collision strengths and transition probabilities for [\ion{Fe}{iii}] are those
from Zhang (\cite{zha96}) and Quinet (\cite{qui96}), respectively.

\section{Morphology of the nebulae}

\begin{figure*}
  \includegraphics[width=17cm]{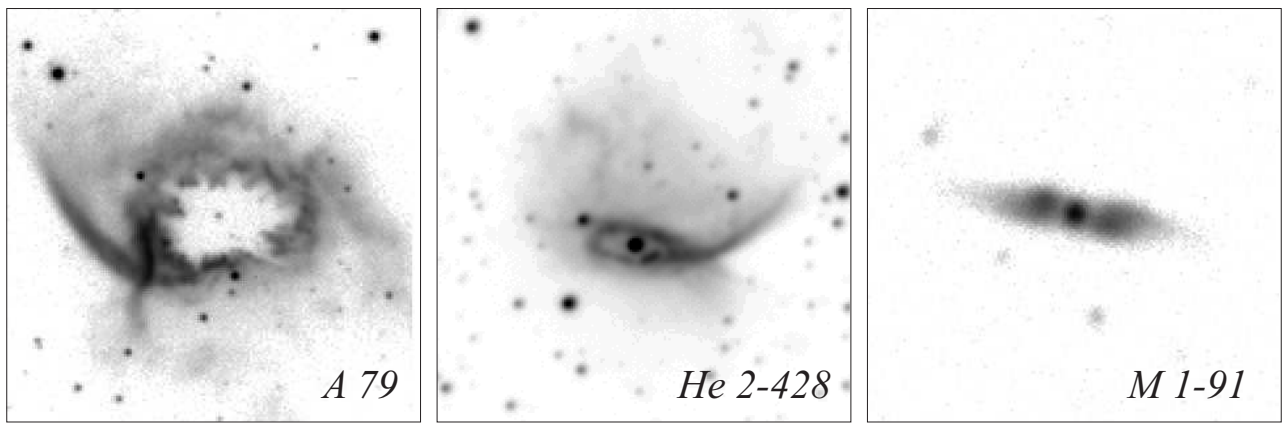}
    \caption{Images of \object{A~79} (in [\ion{N}{ii}], from Balick \cite{B87}),
     \object{He~2-428} (in $\element{H}\alpha+[\ion{N}{II}]$, see text)
     and \object{M~1-91} (in H$\alpha$, Rauch \cite{rau99}),
     in a logarithmic intensity scale.
     The size of each box is 60 arcsec.}
\label{fimag}
\end{figure*}

Figure~\ref{fimag} shows the images of the three nebulae.
In addition to our INT image of \object{He~2-428}, a [\ion{N}{II}] image of
\object{A~79} was kindly provided by B.~Balick (from Balick \cite{B87}) and
an H$\alpha$ one of \object{M~1-91} by T.~Rauch (Rauch \cite{rau99}).

\object{A~79} and \object{He~2-428} (Fig.~\ref{fimag}, left and centre) are
similar, looking like rather evolved bipolar nebulae. They are
composed of an irregular ring from which faint bipolar lobes depart,
which appear to have almost vanished into the interstellar medium. Note
the asymmetry in the surface brightness of the upper lobes, which have
one edge much brighter than the other ones (the NE edge for
\object{A~79} and the NW one for
\object{He~2-428}).

\object{He~2-428} has a bright core inside the ring, while a fainter
star-like object appears roughly at the centre of symmetry of the ring of
\object{A~79}.
\object{M~1-91} (Fig.~\ref{fimag}, right) also has a
bright core, and possesses highly collimated lobes similar to those of
its more studied sibling \object{M~2-9} (cf. Doyle et al.\ \cite{doy00}).

\section{The nuclei of the nebulae}

\begin{figure*}
\centering
  \includegraphics[width=17cm]{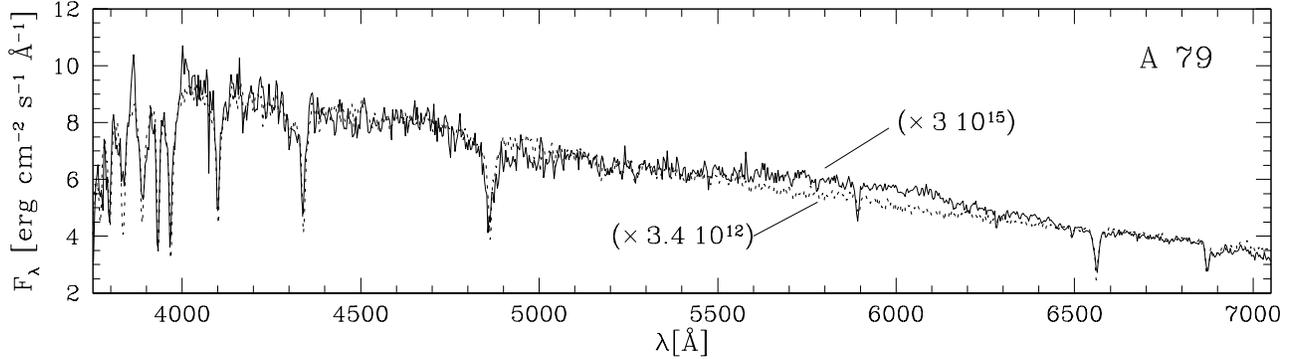}
    \caption{The spectrum of the star at the symmetry centre of \object{A~79}
    (continuous line) compared with an F0~V star spectrum (dotted line) from
    Jacoby et al.\ (\cite{jac84}). Both spectra have been normalized by the
    factors enclosed in parentheses}
    \label{figa79}
\end{figure*}

\begin{figure*}
\centering
  \includegraphics[width=17cm]{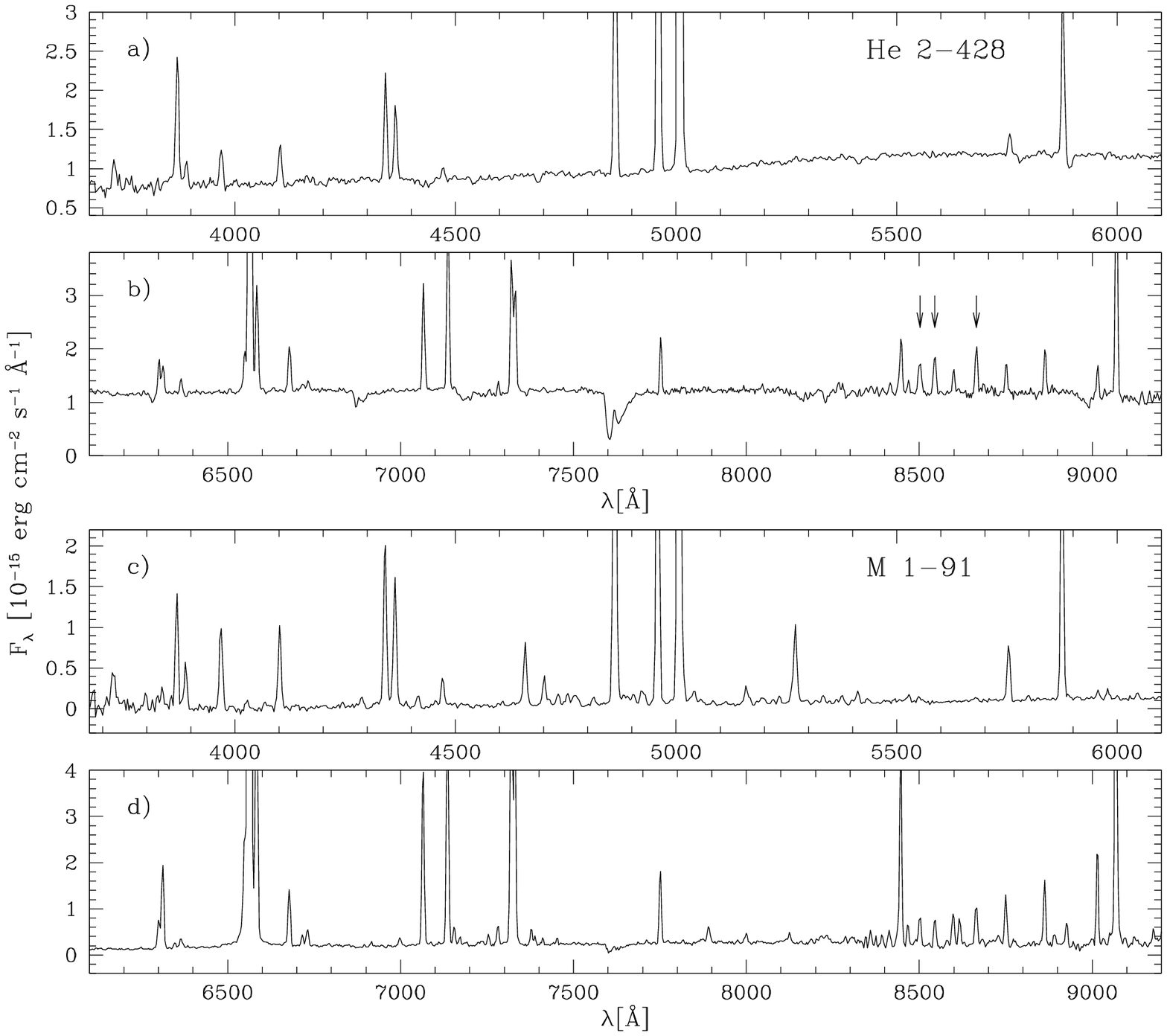}
    \caption{Spectra of the cores of \object{He~2-428} ({\bf a}, {\bf b})
    and \object{M~1-91} ({\bf c}, {\bf d}). The arrows in panel {\bf b} show the
    positions where the \ion{Ca}{ii} triplet lines appear in emission in
    \object{He~2-428}, blended with three Paschen lines}
    \label{figco}
\end{figure*}

The central source in \object{A~79} is found to have a stellar spectra, which
is shown in Fig.~\ref{figa79} and discussed in Sect.~\ref{na79}, but the
nuclei of \object{He~2-428} and \object{M~1-91} have emission line spectra.
Figure~\ref{figco} shows a close-up of the low-resolution spectra of the
bright cores of \object{He~2-428} and \object{M~1-91}; Tables~2 and 3
show the line intensities relative to H$\beta$ before and after the
extinction correction, the intensity measured for H$\beta$, the logarithmic
extinction $c$ and the size of the
region over which the spectra have been extracted.
Although both cores show indications of emission at high densities, as
detailed below, their spectra are quite different.
Both cores show
the emission lines usually found in low-excitation nebulae
(\ion{H}{i}, \ion{He}{i}, [\ion{O}{ii}], [\ion{O}{iii}],
[\ion{N}{ii}], [\ion{S}{ii}], [\ion{Ar}{iii}]).
\object{M~1-91} shows also numerous lines of \ion{Fe}{ii},
[\ion{Fe}{ii}], [\ion{Fe}{iii}], [\ion{Fe}{iv}], [\ion{Ni}{ii}] --
most of them characteristic of high-density emission nebulae.
However, \object{He~2-428} does not have a single \element{Fe} line, but does
show the \element{Ca} IR triplet in emission.
\object{He~2-428} also has a quite strong continuum emission.

\subsection{The central star of \object{A~79}}
\label{na79}

Figure~\ref{figa79} shows the low-resolution spectrum of the stellar-like
object which is located roughly at the centre of symmetry of the nebular
ring of \object{A~79}.  At variance with \object{He~2-428} and
\object{M~1-91},
no nebular component is found in the core, and the spectrum is typical
of a star of intermediate spectral type. This stellar spectrum has
been compared with those in the library of stellar spectra of Jacoby
et al.\ (\cite{jac84}).
The best visual fit is obtained with an F0~V star,
whose spectrum is also shown in Fig.~\ref{figa79}, but our spectrum
does not allow us to exclude the adjacent spectral types or to define
the luminosity class.  The spectrum has been corrected for extinction
using $A_{\rm V}=1.7$, the value leading to the best fit with the F0~V
star spectrum.  This value is equivalent to $c\simeq0.7$, only
somewhat higher than the optical depths found for the extended nebula
of \object{A~79} ($c=0.39\mathrm{\ and\ }0.48$), which supports the
possibility that the star is in fact associated with
the nebula. The agreement between the two spectra breaks in the
range $\lambda\lambda$5500--6400, where the central star of
\object{A~79} shows excess emission, a bump, over the comparison star.
Similar excesses were found in two other field stars observed in the
same exposure.
The flux calibration was checked with another
standard star and with data from different IRAF directories, but the bumps
remained.
They could be due to a deviation of the extinction curve from a linear trend
around this wavelength region, known as the very broad
structure in the extinction curve.
The deviation seems to correlate with reddening
and its origin is not well understood (see e.g.\ Jenniskens
\cite{jen94}).  However, the deviation found for the central star of
\object{A~79}, which amounts to $\sim0.1$~mag, is higher than the
value expected from its reddening, $\sim0.034$~mag.
Therefore, other explanations, including some uncontrolled instrumental effect,
cannot be excluded.

\subsection{The core of \object{He~2-428}}
\label{nhe2428}

\begin{table*}
\caption[ ]{Line intensities in the core of \object{He~2-428}}
\begin{tabular}{lllll|lllll}
\hline
\noalign{\smallskip}
$\lambda_{\rm obs}$ & Identification & $(I_\lambda/I_\beta)_0$ &
	$I_\lambda/I_\beta$&&&
$\lambda_{\rm obs}$ & Identification & $(I_\lambda/I_\beta)_0$ &
	$I_\lambda/I_\beta$ \\
\noalign{\smallskip}
\hline
\noalign{\smallskip}
3726.3 & [\ion{O}{ii}] 3726.0 + 3728.8  & 0.10 & 0.37 
	&&&  6678.8 & \ion{He}{i} 6678.2   & 0.18 & 0.042 \\
3869.4 & [\ion{Ne}{iii}] 3868.8 & 0.36 & 1.14
	&&&  6718.1 & [\ion{S}{ii}] 6716.5 & 0.033 & 0.0077 \\
3889.6 & \ion{He}{i} 3888.7 + H8 3889.1 & 0.073 & 0.22
	&&&  6732.5 & [\ion{S}{ii}] 6730.9 & 0.043 & 0.0098 \\
3968.9 & [\ion{Ne}{iii}] 3967.5 + H$\epsilon$ 3970.1 & 0.11 & 0.30
	&&&  7066.3 & \ion{He}{i} 7065.3   & 0.42 & 0.081 \\
4102.1 & H$\delta$ 4101.7  & 0.12 & 0.28
	&&&  7136.9 & [\ion{Ar}{iii}] 7135.8 & 0.65 & 0.12 \\
4341.4 & H$\gamma$ 4340.5     & 0.27 & 0.49
	&&&  7282.4 & \ion{He}{i} 7281.4   & 0.049 & 0.0084 \\
4364.1 & [\ion{O}{iii}] 4363.2 & 0.20 & 0.35
	&&&  7321.0 & [\ion{O}{ii}] 7319.7 & 0.54 & 0.090 \\
4471.9 & \ion{He}{i} 4471.5   & 0.034 & 0.053 
	&&&  7331.5 & [\ion{O}{ii}] 7330.2 & 0.43 & 0.071 \\
4862.3 & H$\beta$ 4861.2    & 1.00 & 1.00
	&&&  7752.6 & [\ion{Ar}{iii}] 7751.1 & 0.20 & 0.027 \\
4923.6 & \ion{He}{i} 4922.2  & 0.022 & 0.021
	&&&  8416.0 & Pa19 8413.3      & 0.046 & 0.0044  \\
4960.0 & [\ion{O}{iii}] 4958.9 & 2.05 & 1.83
	&&&  8447.5 & \ion{O}{i} 8446.5  & 0.26 & 0.025  \\
5007.9 & [\ion{O}{iii}] 5006.8 & 6.42 & 5.48
	&&&  8469.4 & Pa17 8467.3     & 0.061 & 0.0057  \\
5756.4 & [\ion{N}{ii}] 5754.6 & 0.057 & 0.023
	&&&  8501.5 & \ion{Ca}{ii} 8498.0 + Pa16 8502.5 & 0.16 & 0.015  \\
5877.0 & \ion{He}{i} 5875.7   & 0.41 & 0.17
	&&&  8545.1 & \ion{Ca}{ii} 8542.1 + Pa15 8545.4 & 0.17 & 0.015 \\
6302.3 & [\ion{O}{i}] 6300.3  & 0.16 & 0.047
	&&&  8599.6 & Pa14 8598.4    & 0.095 & 0.0083 \\
6312.0 & [\ion{S}{iii}] 6312.1 & 0.14 & 0.040
	&&&  8665.4 & \ion{Ca}{ii} 8662.1 + Pa13 8665.0 & 0.18 & 0.015 \\
6364.8 & [\ion{O}{i}] 6363.8  & 0.059 & 0.016
	&&&  8751.8 & Pa12 8750.5    & 0.12 & 0.0094 \\
6549.1 & [\ion{N}{ii}] 6548.0 & 0.12 & 0.031
	&&&  8864.1 & Pa11 8862.8   & 0.18 & 0.014 \\
6563.9 & H$\alpha$ 6562.8    & 11.6 & 2.89
	&&&  9070.4 & [\ion{S}{iii}] 9068.9 & 1.17 & 0.101 \\
6584.4 & [\ion{N}{ii}] 6583.4 & 0.42 & 0.10
	&&&  &  & &  \\
\noalign{\smallskip}
\hline
\noalign{\smallskip}
\multicolumn{3}{l}{$I(\mathrm{H}\beta)=4.0\times10^{-14}$~
		$\mathrm{erg}\,\mathrm{cm}^{-2}\,\mathrm{s}^{-1}$}\\
\multicolumn{2}{l}{$c=1.63$}\\
\multicolumn{2}{l}{$Size=3\farcs7$}\\
\noalign{\smallskip}
\hline
\end{tabular}
\end{table*}

The spectrum of the nucleus of \object{He~2-428} is shown in
Figs.~\ref{figco}a and \ref{figco}b; the measured and reddening-corrected
intensities are listed in Table~2.
The logarithmic extinction $c$ has been calculated by fitting the ratios
$I(\mathrm{H}\alpha)/I(\mathrm{H}\beta)$,
$I(\mathrm{H}\gamma)/I(\mathrm{H}\beta)$,
$I(\mathrm{H}\delta)/I(\mathrm{H}\beta)$,
$I(\mathrm{Pa}11)/I(\mathrm{H}\beta)$ and
$I(\mathrm{Pa}12)/I(\mathrm{H}\beta)$ to their recombination values for
suitable physical conditions (see below).  Optical depth effects might
be affecting the \ion{H}{i} intensities, as in \object{M~1-91}
(see Sect.~\ref{nm191}),
but since the reddening-corrected ratios are within 10\% of
their recombination values for $\log(N_\mathrm{e})\sim5.5$ and
$T_\mathrm{e}\sim15\,000\>\mathrm{K}$, these effects, if present, must be
small.

The extinction-corrected line ratios imply densities in the ranges:
$\log(N_{\rm e}[\ion{S}{ii}])=3.2\mbox{--}3.3$,
$\log(N_{\rm e}[\ion{N}{ii}])=5.1\mbox{--}5.5$,
$\log(N_{\rm e}[\ion{O}{iii}])=5.4\mbox{--}6.3$ for
$T_{\rm e}=10\,000\mbox{--}20\,000$~K (with the highest values for
$N_{\rm e}[\ion{N}{ii}]$ and $N_{\rm e}[\ion{O}{iii}]$
corresponding to the lower temperature).

The core shows a relatively strong continuum, with absorption lines from
\ion{He}{II} and some interstellar features.
When corrected for reddening, the continuum and the absorption lines
can be reproduced with an O5 star and an additional continuum component
with an apparent onset at $\sim5000$~\AA.
This additional component could be due to a late-type companion or arise in the
line-emitting gas.
The minimum contribution of this additional component to the observed continuum
can be estimated by assuming that the continuum up to 5000~\AA\ is mainly due
to the hot star.
We find that a minimum of 20--25\% of the observed continuum should be due to
this additional continuum component.
The strongest source of nebular continuum that could produce an observable
onset at 5000~\AA, is the recombination continuum of \ion{H}{i}.
An estimate of its contribution to the observed continuum can be obtained from
the intensity measured for H$\beta$ and the emission coefficients for this line
and the continuum (Osterbrock \cite{ost89}).
The result is that just a few percent of the observed continuum can be of
nebular origin.
Therefore, the additional continuum component is probably due to
a late-type stellar companion of the hot star.

The spectrum in Fig.~\ref{figco} shows also that the \ion{Ca}{ii}
infrared triplet lines appear in emission, blended with three Paschen
lines.
The \ion{Ca}{ii} line intensities can be estimated by subtracting the
expected contribution of the Paschen lines.  This has been done by
linearly interpolating the intensities of the surrounding Pa12,
Pa14 and Pa17 lines.  The three
\ion{Ca}{ii} lines turn out to have very similar intensities, with
$I(\lambda8498)/I(\mathrm{H}\beta)=0.086$,
$I(\lambda8542)/I(\mathrm{H}\beta)=0.078$ and
$I(\lambda8662)/I(\mathrm{H}\beta)=0.062$, i.e. in the ratio
1.0:0.9:0.7, whereas the ratio for optically thin emission is 1:9:5.
A near equality of the \ion{Ca}{ii} intensities is usually found in
all objects where the triplet is in emission and is explained as due
to a high optical depth.  The \ion{Ca}{ii} infrared triplet has been
observed in emission in a variety of objects: some Be stars,
high-luminosity young stellar objects (some with bipolar molecular
outflows), certain active galactic nuclei, cataclysmic variable stars
and T~Tauri stars (Persson \cite{per88a} and references therein),
symbiotic stars (Cieslinski et al.\ \cite{cie97}) and the
proto-planetary nebula \object{He~3-401} (Garc\'\i a-Lario et al.\
\cite{gar99}).
These objects support a dense neutral region in close proximity
to a source of photons that is capable of ionizing \element[+]{Ca} --
whose ionization potential is very low, 11.9~eV -- unless it is well
shielded.  Therefore, for most of these \ion{Ca}{ii} triplet emitters,
the emission is expected to arise in a disk.
Some of the \ion{Ca}{ii} triplet emitters also show emission in
[\ion{Ca}{ii}] $\lambda$7291 and $\lambda$7324, transitions which arise
in the metastable lower levels of the triplet lines.
These forbidden lines are suppressed at high densities, so that their absence
in the core of \object{He~2-428} suggests
$N_{\rm e}>10^{10}\mathrm{\ cm}^{-3}$
in the \ion{Ca}{ii} emitting region (Ferland \& Persson
\cite{fer89}).  Most of the \ion{Ca}{ii} emitters, unlike \object{He~2-428},
also show \ion{Fe}{ii} emission, which at least in active galactic nuclei seems
to be correlated with \ion{Ca}{ii} emission (Persson \cite{per88b}),
although the \ion{Ca}{ii} lines must arise in a more dense and
optically thick material than \ion{Fe}{ii} emission.

\subsection{The core of \object{M~1-91}}
\label{nm191}

\begin{table*}
\caption[ ]{Line intensities in the core of \object{M~1-91}}
\begin{tabular}{lllll|lllll}
\hline
\noalign{\smallskip}
$\lambda_{\rm obs}$ & Identification & $(I_\lambda/I_\beta)_0$ &
	$I_\lambda/I_\beta$ &&&
$\lambda_{\rm obs}$ & Identification & $(I_\lambda/I_\beta)_0$ &
	$I_\lambda/I_\beta$ \\
\noalign{\smallskip}
\hline
\noalign{\smallskip}
3724.5 & [\ion{O}{ii}] 3726.0 + 3728.8 & 0.099 & 0.23 
	&&& 5550.8 & \ion{N}{ii} 5552.0 ? & 0.0088 & 0.0055 \\
3797.3 & H10 3796.9 & 0.024 & 0.053
	&&& 5754.0 & [\ion{N}{ii}] 5754.6 & 0.10 & 0.057 \\
3835.0 & H9 3835.4 & 0.030 & 0.063
	&&& 5798.8 & [\ion{Fe}{iv}] 5798.0 ($^4G\,\mbox{--}^2D$)
		& 0.0088 & 0.0049 \\
3868.3 & [\ion{Ne}{iii}] 3868.8 & 0.19 & 0.39
	&&&   & [\ion{Fe}{iv}] 5800.4 ($^4G\,\mbox{--}^2D$) &  &  \\
3888.1 & \ion{He}{i} 3888.7 + H8 3889.1 & 0.079 & 0.16
	&&& 5875.4 & \ion{He}{i} 5875.7 & 0.50 & 0.27 \\
3968.3 & [\ion{Ne}{iii}] 3967.5 + H$\epsilon$ 3970.1 & 0.16 & 0.30
	&&& 5956.8 & \ion{Si}{ii} 5957.6 & 0.012 & 0.0063 \\
4027.3 & \ion{He}{i} 4026.2   & 0.014 & 0.026 
	&&& 5978.4 & \ion{Si}{ii} 5978.9 & 0.018 & 0.0094 \\
4101.4 & H$\delta$ 4101.7  & 0.15 & 0.26
	&&& 6000.4 & [\ion{Ni}{iii}] 6000.2 (2$F$) & 0.012 & 0.0062 \\
4243.9 & [\ion{Fe}{ii}] 4244.0 (21$F$) & 0.0083 & 0.013
	&&& 6046.1 & \ion{O}{i} 6046.4 & 0.012 & 0.0062 \\
4287.2 & [\ion{Fe}{ii}] 4287.4 (7$F$) & 0.025 & 0.039
	&&& 6300.3 & [\ion{O}{i}] 6300.3  & 0.086 & 0.042 \\
4388.0 & \ion{He}{i} 4387.9   & 0.0064 & 0.0091 
	&&& 6312.0 & [\ion{S}{iii}] 6312.1 & 0.26 & 0.12 \\
4340.3 & H$\gamma$ 4340.5 & 0.31 & 0.45
	&&& 6347.3 & \ion{Si}{ii} 6347.1  & 0.022 & 0.0096 \\
4362.8 & [\ion{O}{iii}] 4363.2 & 0.25 & 0.36
	&&& 6364.8 & [\ion{O}{i}] 6363.8  & 0.041 & 0.018 \\
4415.1 & [\ion{Fe}{ii}] 4416.3 (6$F$) & 0.023 & 0.032
	&&& 6384.3 & \ion{Fe}{ii} 6383.8 (...)  & 0.0044 & 0.0019 \\
4455.0 & [\ion{Fe}{ii}] 4452.1 (7$F$) & 0.0059 & 0.0079
	&&& 6547.3 & [\ion{N}{ii}] 6548.0 & 0.40 & 0.16 \\
       & [\ion{Fe}{ii}] 4458.0 (6$F$) &  & 
	&&& 6564.8 & H$\alpha$ 6562.8  & 18.41 & 7.58 \\
4471.0 & \ion{He}{i} 4471.5   & 0.055 & 0.073 
	&&& 6583.1 & [\ion{N}{ii}] 6583.4 & 0.88 & 0.36 \\
4606.6 & [\ion{Fe}{iii}]~4607.1~(3$F$) & 0.0054 & 0.0065
	&&& 6678.1 & \ion{He}{i} 6678.2 & 0.18 & 0.070 \\
4657.9 & [\ion{Fe}{iii}]~4658.1~(3$F$) & 0.12 & 0.14
	&&& 6716.0 & [\ion{S}{ii}] 6716.5 & 0.036 & 0.014 \\
4701.2 & [\ion{Fe}{iii}] 4701.6 (3$F$) & 0.054 & 0.060
	&&& 6730.7 & [\ion{S}{ii}] 6730.9 & 0.055 & 0.021 \\
4712.2 & \ion{He}{i} 4713.2 & 0.011 & 0.013
	&&& 6915.4 & [\ion{Cr}{iv}] 6914.8 (2$F$) & 0.014 & 0.0052 \\
4732.7 & [\ion{Fe}{iii}] 4733.9 (3$F$) & 0.023 & 0.025
	&&& 6998.3 & [\ion{Fe}{iv}] 6997.1 ($^4P\,\mbox{--}^2D$) & 0.031
		& 0.011 \\
4754.0 & [\ion{Fe}{iii}] 4754.8 (3$F$) & 0.028 & 0.030
	&&& 7065.2 & \ion{He}{i} 7065.3 & 0.55 & 0.19 \\
4771.6 & [\ion{Fe}{iii}] 4769.6 (3$F$) & 0.033 & 0.035
	&&& 7088.0 & [\ion{Fe}{iii}] 7088.5 (15$F$) & 0.011 & 0.0037 \\
       & [\ion{Fe}{iii}] 4777.7 (3$F$) &  & 
	&&& 7135.7 & [\ion{Ar}{iii}] 7135.8 & 0.75 & 0.25 \\
4813.6 & [\ion{Fe}{ii}] 4814.6 (20$F$) & 0.020 & 0.021
	&&& 7155.2 & [\ion{Fe}{ii}] 7155.1 (14$F$) & 0.061 & 0.021 \\
4861.1 & H$\beta$ 4861.2 & 1.00 & 1.00
	&&& 7171.5 & [\ion{Fe}{ii}] 7172.0 (14$F$) & 0.026 & 0.0087 \\
4881.9 & [\ion{Fe}{iii}] 4881.0 (2$F$) & 0.0080 & 0.0078
	&&& 7225.0 & [\ion{Fe}{iv}] 7222.9 ($^4D\,\mbox{--}^4F$) & 0.012
		& 0.0040 \\
4890.5 & [\ion{Fe}{ii}] 4889.6 (4$F$) & 0.0083 & 0.0081
	&&& 7235.4 & \ion{C}{ii} 7236.2 & 0.021 & 0.0070 \\
4905.2 & [\ion{Fe}{ii}] 4905.4 (20$F$) & 0.016 & 0.015
	&&& 7254.2 & \ion{O}{i} 7254.4 & 0.030 & 0.0099 \\
       & [\ion{Fe}{iv}] 4906.7 ($^4G\,\mbox{--}^4F$) &   &  
	&&& 7281.3 & \ion{He}{i} 7281.4 & 0.057 & 0.019  \\
4921.7 & \ion{He}{i} 4921.9   & 0.031 & 0.030 
	&&& 7319.9 & [\ion{O}{ii}] 7319.7 & 0.96 & 0.31  \\
4930.8 & [\ion{Fe}{iii}] 4930.5 (1$F$) & 0.013 & 0.012
	&&& 7330.2 & [\ion{O}{ii}] 7330.2 & 0.75 & 0.24  \\
4958.5 & [\ion{O}{iii}] 4958.9 & 0.80 & 0.74
	&&& 7378.0 & [\ion{Ni}{ii}] 7377.9 (2$F$) & 0.0254 & 0.017 \\
5006.5 & [\ion{O}{iii}] 5006.8 & 2.50 & 2.26
	&&& 7387.8 & [\ion{Fe}{ii}] 7388.2 (14$F$) & 0.029 & 0.0092 \\
5040.3 & \ion{Si}{ii} 5041.0 & 0.025 & 0.022
	&&& 7411.5 & [\ion{Ni}{ii}] 7411.6 (2$F$) & 0.021 & 0.0066 \\
5084.2 & [\ion{Fe}{iii}] 5084.8 (1$F$) & 0.0081 & 0.0069
	&&& 7452.7 & [\ion{Fe}{ii}] 7452.5 (14$F$) & 0.019 & 0.0059  \\
5112.4 & [\ion{Fe}{ii}] 5111.6 (19$F$) & 0.0054 & 0.0045
	&&& 7712.3 & \ion{Fe}{ii} 7711.4 (73) & 0.012 & 0.0034 \\
5159.5 & [\ion{Fe}{ii}] 5157.9 (19$F$) & 0.043 & 0.035
	&&& 7750.9 & [\ion{Ar}{iii}] 7751.1 & 0.22 & 0.062  \\
       & [\ion{Fe}{ii}] 5158.8 (18$F$) &  &  
	&&& 7816.2 & \ion{He}{i} 7816.2 & 0.0095 & 0.0026 \\
5194.9 & [\ion{Ar}{iv}] 5191.8 & 0.022 & 0.018
	&&& 7890.8 & [\ion{Ni}{iii}] 7889.9 (1$F$) & 0.068 & 0.019  \\
       & [\ion{N}{ii}] 5197.9 + 5200.3 &   &  
	&&& 7999.2 & [\ion{Cr}{ii}] 7999.9 (1$F$) & 0.036 & 0.0091  \\
5219.9 & [\ion{Fe}{ii}] 5220.1 (19$F$) & 0.0068 & 0.0053
	&&& 8124.5 & [\ion{Cr}{ii}] 8125.2 (1$F$) & 0.042 & 0.010  \\
5234.1 & [\ion{Fe}{iv}] 5234.2 ($^4G\,\mbox{--}^2F$) & 0.0068 & 0.0053
	&&& 8157.8 & \ion{Fe}{ii} 8157.5 ($e^6D\,\mbox{--}\,5p^4F$) ?
			& 0.0076 & 0.0019  \\
       & \ion{Fe}{ii} 5234.6 (49) &  & 
	&&& 8187.1 & \ion{N}{i} 8188.0 & 0.021 & 0.0051  \\
5270.3 & [\ion{Fe}{ii}] 5261.6 (19$F$) & 0.16 & 0.12
	&&& 8345.1 & \ion{He}{i} 8342.4 + Pa23 8345.6 & 0.028 & 0.0064 \\
       & [\ion{Fe}{iii}] 5270.4 (1$F$) &  & 
	&&& 8359.5 & Pa22 8359.0 + \ion{He}{i} 8361.7 & 0.058 & 0.013 \\
       & [\ion{Fe}{ii}] 5273.4 (18$F$) &  & 
	&&& 8374.3 & Pa21 8374.5 & 0.022 & 0.0050 \\
5317.9 & \ion{Fe}{ii} 5316.6 (49) & 0.0053 & 0.0039
	&&& 8392.0 & Pa20 8392.4 & 0.031 & 0.0069  \\
5333.6 & [\ion{Fe}{ii}] 5333.7 (19$F$) & 0.015 & 0.011
	&&& 8412.9 & Pa19 8413.3 & 0.049 & 0.011  \\
5348.0 & ? & 0.0027 & 0.0019
	&&& 8446.3 & \ion{O}{i} 8446.5 & 0.60 & 0.13  \\
5363.9 & \ion{Fe}{ii} 5362.9 (48) & 0.0063 & 0.0045
	&&& 8467.0 & Pa17 8467.3 & 0.069 & 0.015  \\
5376.2 & [\ion{Fe}{ii}] 5376.5 (19$F$) & 0.014 & 0.0095
	&&& 8501.6 & Pa16 8502.5 & 0.099 & 0.022  \\
5411.8 & [\ion{Fe}{iii}] 5412.0 (1$F$) & 0.021 & 0.014
	&&& 8545.5 & Pa15 8545.4 & 0.086 & 0.018 \\
5433.9 & [\ion{Fe}{ii}] 5433.1 (18$F$) & 0.0074 & 0.0050
	&&& 8580.5 & [\ion{Cl}{ii}] 8578.7 + \ion{He}{i} 8582.5 &
					0.028 & 0.0060 \\
5526.5 & [\ion{Fe}{ii}] 5527.3 (17$F$) & 0.015 & 0.0096
	&&& 8598.2 & Pa14 8598.4 & 0.11 & 0.024 \\
\noalign{\smallskip}
\hline
\end{tabular}
\end{table*}
\setcounter{table}{2}
\begin{table*}
\caption[ ]{{\em continued}}
\begin{tabular}{lllll|lllll}
\hline
\noalign{\smallskip}
$\lambda_{\rm obs}$ & Identification & $(I_\lambda/I_\beta)_0$ &
	$I_\lambda/I_\beta$ &&&
$\lambda_{\rm obs}$ & Identification & $(I_\lambda/I_\beta)_0$ &
	$I_\lambda/I_\beta$ \\
\noalign{\smallskip}
\hline
\noalign{\smallskip}
8616.6 & [\ion{Fe}{ii}] 8617.0 (13$F$) & 0.086 & 0.018
	&&& 8862.7 & Pa11 8862.8 & 0.21 & 0.041 \\
8664.7 & Pa13 8665.0 & 0.13 & 0.027
        &&& 8891.5 & [\ion{Fe}{ii}] 8891.9 (13$F$) & 0.039 & 0.0076 \\
8682.4 & \ion{N}{i} 8680.2 + 8683.4 & 0.022 & 0.0046
	&&& 8926.0 & \ion{Fe}{ii} 8926.7 ($e^4D\,\mbox{--}\,5p^4D$)
		& 0.069 & 0.013 \\
8728.6 & [\ion{Fe}{iii}] 8728.9 (8$F$) & 0.034 & 0.0070
	&&& 9014.8 & Pa10 9014.9 & 0.27 & 0.051 \\
8750.3 & Pa12 8750.5 & 0.16 & 0.032
	&&& 9068.7 & [\ion{S}{iii}] 9068.9 & 1.53 & 0.29 \\
8840.0 & [\ion{Fe}{iii}] 8838.2 (8$F$) & 0.034 & 0.0070
	&&&  &  &  &  \\
\noalign{\smallskip}
\hline
\noalign{\smallskip}
\multicolumn{3}{l}{$I(\mathrm{H}\beta)=5.7\times10^{-14}$~
		$\mathrm{erg}\,\mathrm{cm}^{-2}\,\mathrm{s}^{-1}$}\\
\multicolumn{2}{l}{$c=1.04$}\\
\multicolumn{2}{l}{$Size=4\farcs1$}\\
\noalign{\smallskip}
\hline
\end{tabular}
\end{table*}

The spectrum of the core of \object{M~1-91} is shown in Figs.~\ref{figco}c and
\ref{figco}d; the measured and reddening-corrected intensities are listed in
Table~3.
The relative intensities of the \ion{H}{i} lines do not agree with case B
expectations for any amount of reddening, implying that the core of
\object{M~1-91} must be optically thick in H$\alpha$ (Drake \& Ulrich
\cite{dra80}).
Since the amount of extinction has been estimated by averaging those
derived for the outer regions of the nebulae (see Sect.~4), and it is
probably higher in the core, the reddening-corrected intensities are
mainly given for illustrative purposes.

Line identifications rely on the comparison of the spectrum with those
of \object{He~2-25} and \object{Th~2-B} (Corradi
\cite{cor95}), \object{M8} (Esteban et al.\ \cite{est99}),
\object{$\eta$~Car} (Hamann et al.\ \cite{ham94}), \object{RR~Tel}
(McKenna et al.\ \cite{mck97}), \object{IC~4997} (Hyung et al.\
\cite{hyu94}), the peculiar B[e] star \object{HD~45677} (Swings
\cite{swi73}), and the tables by Moore (\cite{moo45}).
The lines can have contributions of fainter transitions, mainly from
[\ion{Fe}{ii}] and \ion{Fe}{ii}, and some weak lines of \ion{Mg}{ii}
may also be present.  The spectrum is very similar to those found in
the cores of its sibling
\object{M2-9}  (as noted previously by Calvet \& Cohen \cite{cal78} and
Goodrich \cite{goo91}) and other highly collimated planetary nebulae:
\object{He~2-25} and \object{Th~2-B} (Corradi \cite{cor95}).
The four objects have however different degrees of excitation and the
intensities of the [\ion{Fe}{iv}], [\ion{Fe}{iii}], [\ion{Fe}{ii}] and
\ion{Fe}{ii} lines are also different, probably reflecting different
conditions of density, and maybe, different evolutionary states.

The extinction-corrected line ratios, after allowing for a possibly
somewhat higher extinction correction, imply densities in the ranges:
$\log(N_{\rm e}[\ion{S}{ii}])=3.55\pm0.10$, $\log(N_{\rm
e}[\ion{N}{ii}])=5.15\pm0.25$, $\log(N_{\rm
e}[\ion{O}{iii}])=6.4\pm0.4$, $\log(N_{\rm e}[\ion{Fe}{ii}])\sim6$ and
$\log(N_{\rm e}[\ion{Fe}{iii}])=6.5\pm0.5$ for $T_{\rm
e}=10\,000\mbox{--}20\,000$~K.

\section{Chemical abundances in the lobes or rings}
\label{abund}

\begin{table*}
\caption[ ]{Line intensities in the ring of \object{A~79}}
\begin{tabular}{lllllllll}
\hline
\noalign{\smallskip}
\multicolumn{3}{c}{\object{A~79}~A}&\multicolumn{1}{c}{}&
		\multicolumn{3}{c}{\object{A~79}~B}\\
\noalign{\smallskip}
\cline{1-3}\cline{5-7}
\noalign{\smallskip}
$\lambda_{\rm obs}$ & $(I_\lambda/I_\beta)_0$ & $I_\lambda/I_\beta$ &&
	$\lambda_{\rm obs}$ & $(I_\lambda/I_\beta)_0$ & $I_\lambda/I_\beta$
	&& Identification \\
\noalign{\smallskip}
\hline
\noalign{\smallskip}
3727.6 & 2.27 & 3.32 && 3727.1 & 3.02 & 4.12 && [\ion{O}{ii}] 3726.0+3728.8\\
3868.2 & 1.38 & 1.94 && 3867.8 & 1.27 & 1.67 && [\ion{Ne}{iii}] 3868.8\\
4339.8 & 0.41 & 0.49 && 4340.6 & 0.48 & 0.55 && H$\gamma$ 4340.5\\
4685.0 & 1.06 & 1.12 && 4685.1 & 0.84 & 0.88 && \ion{He}{ii} 4685.7\\
4860.8 & 1.00 & 1.00 && 4860.7 & 1.00 & 1.00 && H$\beta$ 4861.2\\
4958.4 & 3.54 & 3.42 && 4958.5 & 2.56 & 2.50 && [\ion{O}{iii}] 4958.9\\
5006.2 & 10.75 & 10.24 && 5006.5 & 8.14 & 7.83 && [\ion{O}{iii}] 5006.8\\
5198.0 & 0.23 & 0.20 && 5199.0 & 0.26 & 0.24 && [\ion{N}{i}] 5197.9+5200.3\\
5753.2 & 0.46 & 0.35 && 5754.6 & 0.45 & 0.36 && [\ion{N}{ii}] 5754.6\\
5875.8 & 0.15 & 0.11 && 5876.3 & 0.23 & 0.18 && \ion{He}{i} 5875.7\\
6300.1 & 0.42 & 0.29 && 6300.6 & 0.58 & 0.43 && [\ion{O}{i}] 6300.3\\
6311.5 & 0.17 & 0.12 && 6312.2 & 0.08 & 0.06 && [\ion{S}{iii}] 6312.1\\
6547.3 & 6.43 & 4.27 && 6547.9 & 7.69 & 5.52 && [\ion{N}{ii}] 6548.0\\
6562.1 & 4.24 & 2.81 && 6562.7 & 3.91 & 2.80 && H$\alpha$ 6562.8\\
6582.6 & 19.38 & 12.78 && 6583.1 & 22.36 & 15.98 && [\ion{N}{ii}] 6583.4\\
6715.7 & 1.43 & 0.92 && 6716.3 & 1.73 & 1.22 && [\ion{S}{ii}] 6716.5\\
6730.1 & 1.09 & 0.70 && 6730.6 & 1.37 & 0.97 && [\ion{S}{ii}] 6730.9\\
\noalign{\smallskip}
\hline
\noalign{\smallskip}
\multicolumn{1}{l}{$I(\mathrm{H}\beta)$}
	&\multicolumn{4}{l}{$1.9\times10^{-15}$~
		$\mathrm{erg}\,\mathrm{cm}^{-2}\,\mathrm{s}^{-1}$}
	&\multicolumn{4}{l}{$2.2\times10^{-15}$~
		$\mathrm{erg}\,\mathrm{cm}^{-2}\,\mathrm{s}^{-1}$}\\
\multicolumn{1}{l}{$c$}&\multicolumn{2}{l}{0.48}
	&\multicolumn{2}{l}{}&\multicolumn{2}{l}{0.39}\\
\multicolumn{1}{l}{$Size$}&\multicolumn{2}{l}{11\farcs8}
	&\multicolumn{2}{l}{}&\multicolumn{2}{l}{8\farcs4}\\
\multicolumn{1}{l}{$d^{\rm a}$}&\multicolumn{2}{l}{18\farcs5}
	&\multicolumn{2}{l}{}&\multicolumn{2}{l}{24\farcs4}\\
\noalign{\smallskip}
\hline
\noalign{\smallskip}
\multicolumn{7}{l}{$^{\rm a}$ Distance from the center of the extracted
	region to the central star}\\
\end{tabular}
\end{table*}

\begin{table*}
\caption[ ]{Line intensities in the ring of \object{He~2-428}}
\begin{tabular}{lllllllll}
\hline
\noalign{\smallskip}
\multicolumn{3}{c}{\object{He~2-428}~A}&\multicolumn{1}{c}{}&
	\multicolumn{3}{c}{\object{He~2-428}~B}\\
\noalign{\smallskip}
\cline{1-3}\cline{5-7}
\noalign{\smallskip}
$\lambda_{\rm obs}$ & $(I_\lambda/I_\beta)_0$ & $I_\lambda/I_\beta$ &&
	$\lambda_{\rm obs}$ & $(I_\lambda/I_\beta)_0$ & $I_\lambda/I_\beta$
	&& Identification \\
\noalign{\smallskip}
\hline
\noalign{\smallskip}
3727.1 & 0.70 & 1.97 && 3727.6 & 0.63 & 1.95 && [\ion{O}{ii}] 3726.0+3728.8\\
4101.9 & 0.18 & 0.36 && 4102.3 & 0.11 & 0.24 && H$\delta$ 4101.7 \\
4340.4 & 0.32 & 0.51 && 4341.2 & 0.36 & 0.60 && H$\gamma$ 4340.5 \\
4861.9 & 1.00 & 1.00 && 4862.2 & 1.00 & 1.00 && H$\beta$ 4861.2 \\
4959.7 & 1.34 & 1.23 && 4959.8 & 1.29 & 1.17 && [\ion{O}{iii}] 4958.9 \\
5007.7 & 4.15 & 3.66 && 5007.9 & 3.92 & 3.42 && [\ion{O}{iii}] 5006.8 \\
5757.2 & 0.05 & 0.02 && 5753.2 & 0.06 & 0.03 && [\ion{N}{ii}] 5754.6 \\
5876.7 & 0.34 & 0.16 && 5876.8 & 0.34 & 0.15 && \ion{He}{i} 5875.7 \\
6300.7 & 0.09 & 0.03 && 6302.0 & 0.15 & 0.05 && [\ion{O}{i}] 6300.3 \\
6549.0 & 0.75 & 0.25 && 6549.2 & 1.11 & 0.34 && [\ion{N}{ii}] 6548.0 \\
6563.8 & 8.38 & 2.77 && 6564.0 & 9.32 & 2.80 && H$\alpha$ 6562.8 \\
6584.3 & 2.22 & 0.73 && 6584.6 & 3.34 & 0.99 && [\ion{N}{ii}] 6583.4 \\
6678.9 & 0.15 & 0.05 && 6679.5 & 0.14 & 0.04 && \ion{He}{i} 6678.2 \\
6716.8 & 0.25 & 0.08 && 6717.8 & 0.36 & 0.10 && [\ion{S}{ii}] 6716.5 \\
6731.2 & 0.28 & 0.09 && 6732.2 & 0.42 & 0.12 && [\ion{S}{ii}] 6730.9 \\
7066.6 & 0.17 & 0.05 && 7066.4 & 0.14 & 0.03 && \ion{He}{i} 7065.3 \\
7137.2 & 0.36 & 0.09 && 7137.1 & 0.41 & 0.10 && [\ion{Ar}{iii}] 7135.8 \\
7320.8 & 0.20 & 0.05 && 7321.6 & 0.30 & 0.06 && [\ion{O}{ii}] 7319.7 \\
7331.3 & 0.16 & 0.04 && 7332.1 & 0.18 & 0.04 && [\ion{O}{ii}] 7330.2 \\
9070.4 & 1.44 & 0.18 && 9070.8 & 1.58 & 0.16 && [\ion{S}{iii}] 9068.9 \\
\noalign{\smallskip}
\hline
\noalign{\smallskip}
\multicolumn{1}{l}{$I(\mathrm{H}\beta)$}
	&\multicolumn{4}{l}{$6.3\times10^{-15}$~
		$\mathrm{erg}\,\mathrm{cm}^{-2}\,\mathrm{s}^{-1}$}
	&\multicolumn{4}{l}{$6.8\times10^{-15}$~
		$\mathrm{erg}\,\mathrm{cm}^{-2}\,\mathrm{s}^{-1}$}\\
\multicolumn{1}{l}{$c$}&\multicolumn{2}{l}{1.30}
	&\multicolumn{2}{l}{}&\multicolumn{2}{l}{1.41}\\
\multicolumn{1}{l}{$Size$}&\multicolumn{2}{l}{3\farcs3}
	&\multicolumn{2}{l}{}&\multicolumn{2}{l}{3\farcs3}\\
\multicolumn{1}{l}{$d^{\rm a}$}&\multicolumn{2}{l}{3\farcs7}
	&\multicolumn{2}{l}{}&\multicolumn{2}{l}{3\farcs7}\\
\noalign{\smallskip}
\hline
\noalign{\smallskip}
\multicolumn{7}{l}{$^{\rm a}$ Distance from the center of the extracted
	region to the core}\\
\end{tabular}
\end{table*}

\begin{table*}
\caption[ ]{Line intensities in the lobes of \object{M~1-91}}
\begin{tabular}{lllllllll}
\hline
\noalign{\smallskip}
\multicolumn{3}{c}{\object{M~1-91}~A}&\multicolumn{1}{c}{}&
	\multicolumn{3}{c}{\object{M~1-91}~B}\\
\noalign{\smallskip}
\cline{1-3}\cline{5-7}
\noalign{\smallskip}
$\lambda_{\rm obs}$ & $(I_\lambda/I_\beta)_0$ & $I_\lambda/I_\beta$ &&
	$\lambda_{\rm obs}$ & $(I_\lambda/I_\beta)_0$ & $I_\lambda/I_\beta$
	&& Identification \\
\noalign{\smallskip}
\hline
\noalign{\smallskip}
3726.7 & 0.90 & 1.95 && 3726.4 & 0.86 & 2.05 && [\ion{O}{ii}] 3726.0
							+ 3728.8\\
3868.5 & 0.039 & 0.076 && 3867.8 & 0.029 & 0.062 && [\ion{Ne}{iii}] 3868.8\\
4101.3 & 0.15 & 0.25 && 4101.3 & 0.16 & 0.29 && H$\delta$ 4101.7 \\
4340.2 & 0.33 & 0.47 && 4340.0 & 0.33 & 0.49 && H$\gamma$ 4340.5 \\
4362.8 & 0.032 & 0.045 && 4362.2 & 0.012 & 0.018 && [\ion{O}{iii}] 4363.2 \\
4471.0 & 0.043 & 0.056 && 4470.6 & 0.031 & 0.042 && \ion{He}{i} 4471.5\\
4657.1 & 0.023 & 0.027 && 4657.5 & 0.020 & 0.023 && [\ion{Fe}{iii}]
							4658.1 (3$F$)\\
4861.0 & 1.00 & 1.00 && 4860.8 & 1.00 & 1.00 && H$\beta$ 4861.2\\
4881.8 & 0.0095 & 0.0094 && 4882.3 & 0.0073 & 0.0072 && [\ion{Fe}{iii}]
							4881.0 (2$F$)\\
4922.0 & 0.017 & 0.017 && 4921.9 & 0.018 & 0.017 && \ion{He}{i} 4921.9\\
4958.6 & 0.54 & 0.51 && 4958.4 & 0.56 & 0.52 && [\ion{O}{iii}] 4958.9\\
4985.6 & 0.0061 & 0.0057 && 4984.1 & 0.016 & 0.015
			&& [\ion{Fe}{iii}] 4985.9 + 4987.2 (2$F$)\\
5006.6 & 1.69 & 1.53 && 5006.3 & 1.73 & 1.55 && [\ion{O}{iii}] 5006.8\\
5198.8 & 0.030 & 0.024 && 5198.0 & 0.030 & 0.023
			&& [\ion{N}{i}] 5197.9 + 5200.3\\
5269.4 & 0.030 & 0.023 && 5270.0 & 0.025 & 0.018 && [\ion{Fe}{iii}]
							5270.4 (1$F$)\\
5753.9 & 0.037 & 0.022 && 5753.9 & 0.039 & 0.022 && [\ion{N}{ii}] 5754.6\\
5875.6 & 0.27 & 0.15 && 5875.2 & 0.27 & 0.14 && \ion{He}{i} 5875.7\\
6300.0 & 0.074 & 0.035 && 6299.7 & 0.070 & 0.030 && [\ion{O}{i}] 6300.3\\
6312.0 & 0.022 & 0.011 && 6311.7 & 0.023 & 0.010 && [\ion{S}{iii}] 6312.1\\
6364.0 & 0.023 & 0.011 && 6363.9 & 0.031 & 0.013 && [\ion{O}{i}] 6363.8\\
6547.7 & 1.57 & 0.69 && 6547.6 & 1.84 & 0.73 && [\ion{N}{ii}] 6548.0\\
6562.7 & 6.54 & 2.87 && 6562.6 & 7.30 & 2.87 && H$\alpha$ 6562.8\\
6583.1 & 4.90 & 2.13 && 6582.6 & 5.73 & 2.23 && [\ion{N}{ii}] 6583.4\\
6677.6 & 0.086 & 0.037 && 6677.3 & 0.098 & 0.037 && \ion{He}{i} 6678.2\\
6716.2 & 0.32 & 0.13 && 6715.8 & 0.35 & 0.13 && [\ion{S}{ii}] 6716.5\\
6730.6 & 0.45 & 0.19 && 6730.2 & 0.51 & 0.19 && [\ion{S}{ii}] 6730.9\\
7064.9 & 0.092 & 0.035 && 7064.4 & 0.11 & 0.036 && \ion{He}{i} 7065.3\\
7135.7 & 0.30 & 0.11 && 7135.1 & 0.35 & 0.11 && [\ion{Ar}{iii}] 7135.8\\
7281.7 & 0.019 & 0.0067 && 7280.3 & 0.025 & 0.0078 && \ion{He}{i} 7281.4\\
7320.0 & 0.096 & 0.033 && 7318.7 & 0.11 & 0.032 && [\ion{O}{ii}] 7319.7\\
7330.0 & 0.074 & 0.026 && 7329.2 & 0.097 & 0.029 && [\ion{O}{ii}] 7330.2\\
7750.6 & 0.088 & 0.027 && 7750.3 & 0.10 & 0.027 && [\ion{Ar}{iii}] 7751.1\\
8446.6 & 0.094 & 0.023 && 8446.3 & 0.068 & 0.014 && \ion{O}{i} 8446.5\\
8545.4 & 0.021 & 0.0051 && 8543.3 & 0.045 & 0.0090 && Pa15 8545.4\\
8597.8 & 0.035 & 0.0083 && 8597.7 & 0.041 & 0.0079 && Pa14 8598.4\\
8664.4 & 0.047 & 0.011 && 8664.6 & 0.052 & 0.010 && Pa13 8665.0\\
8750.6 & 0.043 & 0.0098 && 8750.4 & 0.051 & 0.0095 && Pa12 8750.5\\
8862.1 & 0.058 & 0.013 && 8861.3 & 0.056 & 0.010 && Pa11 8862.8\\
9014.2 & 0.075 & 0.016 && 9014.6 & 0.12 & 0.020 && Pa10 9014.9\\
9068.7 & 1.16 & 0.24 && 9068.3 & 1.39 & 0.24 && [\ion{S}{iii}] 9068.9\\
\noalign{\smallskip}
\hline
\noalign{\smallskip}
\multicolumn{1}{l}{$I(\mathrm{H}\beta)$}
	&\multicolumn{4}{l}{$3.25\times10^{-14}$~
		$\mathrm{erg}\,\mathrm{cm}^{-2}\,\mathrm{s}^{-1}$}
	&\multicolumn{4}{l}{$3.47\times10^{-14}$~
		$\mathrm{erg}\,\mathrm{cm}^{-2}\,\mathrm{s}^{-1}$}\\
\multicolumn{1}{l}{$c$}&\multicolumn{2}{l}{0.97}
	&\multicolumn{2}{l}{}&\multicolumn{2}{l}{1.10}\\
\multicolumn{1}{l}{$Size$}&\multicolumn{2}{l}{4\farcs1}
	&\multicolumn{2}{l}{}&\multicolumn{2}{l}{4\farcs1}\\
\multicolumn{1}{l}{$d^{\rm a}$}&\multicolumn{2}{l}{6\farcs2}
	&\multicolumn{2}{l}{}&\multicolumn{2}{l}{6\farcs2}\\
\noalign{\smallskip}
\hline
\noalign{\smallskip}
\multicolumn{7}{l}{$^{\rm a}$ Distance from the center of the extracted
	region to the core}\\
\end{tabular}
\end{table*}

The measured line intensities for the extended areas (lobes or rings) of the
three objects, along with the values derived for the logarithmic extinction $c$
and the reddening-corrected relative intensities for all lines of interest are
presented in Tables~4, 5 and 6.

\begin{table*}
\caption[ ]{Physical conditions and ionic abundances ($X^{i+}/\mathrm{H}^+$)}
\begin{tabular}{lllllllllllll}
\hline
\noalign{\smallskip}
Position & $N_\mathrm{e}[\ion{S}{ii}]$ & $T_\mathrm{e}[\ion{N}{ii}]$ &
	O$^+$ & O$^{++}$ & N$^+$ & S$^+$ & S$^{++}$ & Ar$^{++}$ &
		Ne$^{++}$ & He$^+$ & He$^{++}$ & Fe$^{++}$\\
 & (cm$^{-3}$) & (K) & &&&& &&&&&\\
\noalign{\smallskip}
\hline
\noalign{\smallskip}
\object{He~2-428}~A & 1100 & 13700 &
	2.7e-5 & 4.9e-5 & 6.9e-6 & 2.4e-7 & 2.6e-6 & 4.3e-7
			& \dots & 0.13 & \dots & \dots \\
\object{He~2-428}~B & 1200 & 14500 &
	2.2e-5 & 3.9e-5 & 8.4e-6 & 2.8e-7 & 2.1e-6 & 4.3e-7
			& \dots & 0.12 & \dots & \dots \\
\object{M~1-91}~A & 2400 & 8700 &
	2.0e-4 & 8.8e-5 & 6.3e-5 & 1.5e-6 & 8.1e-6 & 1.4e-6 &
			1.5e-5 & 0.11 & \dots & 2.3e-6 \\
\object{M~1-91}~B & 2300 & 8600 &
	2.2e-4 & 9.4e-5 & 6.8-5 & 1.6e-6 & 8.3e-6 & 1.5e-6 &
			1.3e-5 & 0.11 & \dots & 1.6e-6 \\
\object{A~79}~A & 110 & 14000 &
	3.7e-5 & 1.4e-4 & 1.2e-4 & 1.9e-6 & 8.8e-6 & \dots
			& 6.9e-5 & 0.095 & 0.102 & \dots \\
\object{A~79}~B & 170 & 12600 &
	6.7e-5 & 1.4e-4 & 1.9e-4 & 3.2e-6 & 6.2e-6 & \dots
			& 8.2e-5 & 0.15 & 0.078 &\dots \\
\noalign{\smallskip}
\hline
\end{tabular}
\end{table*}

\begin{table*}
\caption[ ]{Total abundances,
	[$X/\element{H}]=\log(X/\element{H})+12$}
\begin{tabular}{llllllll}
\hline
\noalign{\smallskip}
Position & [O/H] & log(N/O) & [N/H] & [S/H] & [Ar/H] & [Ne/H] & He/H\\
\noalign{\smallskip}
\hline
\noalign{\smallskip}
\object{He~2-428}~A & 7.88 & $-$0.59& 7.30 & 6.51 & 5.90 &\dots& $\ge0.13$\\
\object{He~2-428}~B & 7.79 & $-$0.42& 7.36 & 6.38 & 5.90 &\dots& $\ge0.12$\\
\object{M~1-91}~A   & 8.46 & $-$0.50& 7.97 & 6.98 & 6.41 & 7.69 & $\ge0.11$\\
\object{M~1-91}~B   & 8.49 & $-$0.51& 7.98 & 7.00 & 6.45 & 7.63 & $\ge0.11$\\
\object{A~79}~A     & 8.45 & $+$0.50 & 8.94 & 7.18 &\dots & 8.15 & 0.19 \\
\object{A~79}~B     & 8.43 & $+$0.45 & 8.89 & 7.04 &\dots & 8.20 & 0.21 \\
\noalign{\smallskip}
\hline
\noalign{\smallskip}
BPN$^{\rm a}$ & 7.95 to 8.87 & $-$0.68 to $+$0.47 & 7.95 to 8.86 & 
6.26 to 7.18 &	6.04 to 6.76 & 7.78 to 8.34 & 0.11 to 0.25 \\
\ion{H}{ii}$^{\rm b}$ & 8.45$\pm0.10$ & $-$0.75$\pm0.10$ & 7.70$\pm0.10$ &
	7.00$\pm0.10$ & 6.40$\pm0.10$ & 7.60$\pm0.10$ & $\ge0.093$ \\
\object{M2-9}$^{\rm c}$ & 8.36 & $-$0.57& 7.79 &\dots&\dots& 7.41 &
	$\ge0.10$\\
\noalign{\smallskip}
\hline
\end{tabular}
\begin{list}{}{}
\item[$^{\rm a}$] A sample of 15 bipolar planetary nebulae
	(Perinotto \& Corradi \cite{per98}; Pottasch et al.\ \cite{pot00})
\item[$^{\rm b}$] \ion{H}{ii} regions in the solar neighbourhood
	(Osterbrock et al.\ \cite{ost92}; Rodr\'\i guez \cite{rod99})
\item[$^{\rm c}$] Derived from the line intensities presented by Barker
	(\cite{bar78})
\end{list}
\end{table*}

The values for the temperatures $T_\mathrm{e}[\ion{N}{ii}]$, densities
$N_\mathrm{e}[\ion{S}{ii}]$ and ionic abundances for the three
nebulae, derived with the atomic data referenced in Sect.~\ref{atdata},
are listed in Table~7.
The [\ion{O}{iii}]~$\lambda4363$ line could be measured in both positions of
\object{M~1-91}, but it has a poor signal-to-noise ratio, it could furthermore
be contaminated by reflection from the core, and leads to temperatures up to
18500~K, which are much higher than the temperatures derived from the
[\ion{N}{ii}] lines (8600 and 8700~K) and have not been used in the abundance
determination.

In \object{A~79}, where the intensity of \ion{He}{ii}~$\lambda4686$ is
similar to that of H$\beta$, the \ion{H} Balmer intensities in Table~4
include the contribution of the lines from the \ion{He} Pickering series.
This contribution was estimated using
$I(\ion{He}{ii}~\lambda4861)=0.05\,I(\ion{He}{ii}~\lambda4686)$ and
$I(\ion{He}{ii}~\lambda6563)=0.134\,I(\ion{He}{ii}~\lambda4686)$
(Storey \& Hummer \cite{sto95}).  The line intensities used in the abundance
determination for \object{A~79} were consequently corrected by multiplying the
line intensities in Table~4 by 1.06 (\object{A~79}~A) and 1.05
(\object{A~79}~B).

The total abundances were derived by estimating the contribution of the
unobserved ions using the relations of Kingsburgh \& Barlow (\cite{kin94});
the results are presented in Table~8.
An estimate of the \element{Fe} abundance for \object{M~1-91} can be obtained
from $\element{Fe}/\element{O}\simeq\element[++]{Fe}/\element[+]{O}\simeq0.01$,
or about one fourth of the solar value.

The derived abundances can be compared with those found for 15 bipolar PNe by
Perinotto \& Corradi (\cite{per98}) and Pottasch et al.\ (\cite{pot00}),
and with the abundances measured in
\ion{H}{ii} regions in the solar neighbourhood (Osterbrock \cite{ost92};
Rodr\'\i guez \cite{rod99}), also shown in Table~8.
\object{A~79} is among the richest PNe in \element{N} and \element{He},
and shows a \element{Ne} overabundance that could be real (Perinotto \&
Corradi \cite{per98}).
\object{M~1-91} and \object{He~2-428} have a $\element{N}/\element{O}$
abundance ratio one order of magnitude lower, unusual but not unique
among bipolar PNe (cf. Fig.~8 in Perinotto \& Corradi \cite{per98}).
The \element{O}, \element{S} and \element{Ar} abundances in
\object{A~79} and \object{M~1-91}, and the \element{Ne} abundance in
\object{M~1-91}, are essentially identical to those measured in nearby
Galactic \ion{H}{ii} regions.  This is not unexpected since the
Galactocentric distances of the three PNe considered here are most
probably between 6 and 10~kpc (this can be deduced from their
distances: see Sect.~\ref{disc}), the same distance range covered by
the \ion{H}{ii} regions whose
abundances are given in Table~8.  However, \object{He~2-428} shows
quite low abundances for most elements relative to \element{H}, when
compared both with the \ion{H}{ii} regions and with the sample of
bipolar PNe.
In fact, the abundances are similar to those found for PNe belonging
to the halo (Howard et al.\ \cite{how97}) and close to the abundances
of some PNe classified as Peimbert's (\cite{pei78}) type~III, i.e
belonging to an intermediate type between disc and halo nebulae
(Maciel et al.\ \cite{mac90}).  This low metallicity suggests that the
central star of \object{He~2-428} had a low-mass progenitor.  The low
upper limit derived for \element{H}$_2$ emission in this nebula by
Guerrero et al.\ (\cite{gue00}) strengthens this conclusion, since
\element{H}$_2$ emission in PNe seems to show some correlation with the mass
of the progenitor star (Zuckerman \& Gatley \cite{zuc88}).

\section{Kinematics of \object{A~79} and \object{He~2-428}}

The expansion velocities of the rings of \object{A~79} and
\object{He~2-428} have been derived by assuming that the rings are
intrinsically circular so that their inclination $i$, or angle between
the polar axis of the ring and the line of sight, can be estimated
from the relative lengths of the major and minor semi-axis in the
[\ion{N}{II}] images.  This inclination angle can then be used to
deproject the velocities measured in the high-resolution spectra.

The ring of \object{A~79}, which is fragmented and broadened, is
estimated to have its major semi-axis, equivalent to the ring radius,
at a position angle $\mathrm{PA}=105\degr\pm5\degr$ with size
$a\simeq11\farcs6$ and inclination $i=49\degr$.  The ring of
\object{He~2-428} is more clearly defined, with its major semi-axis at
$\mathrm{PA}=82\degr$, with $a=6\farcs3$ and $i=68\degr$.

The velocities in the rings and lobes of the nebulae were measured by
fitting Gaussian profiles to the [\ion{N}{II}]~$\lambda$6583 line in
the echelle spectra; this line was chosen since it has a lower thermal
broadening than the H$\alpha$ line.

The slit orientations for \object{A~79} were located $\sim20\degr$ off
the major and minor axes of the ellipse that fits the ring in the image.
The differences in velocity between opposite regions of the ring were
found to be $\Delta V=18$~\kms, for the position near the minor axis
($\mathrm{PA}=33\degr$), and $\Delta V=11$~\kms\ for the position near
the major axis ($\mathrm{PA}=-57\degr$).  These velocities can be
deprojected to the plane of the ring by using the expression $\Delta
V/2=V_{\rm exp}\,\sin\varphi\,\sin i$, where
$\tan\varphi=\tan\vartheta/\cos i$, and $\vartheta$ is the offset
angle between the slit and the major axis of the ellipse.  The
deprojected velocities are consistent, within the errors, with the
hypothesis of a circular ring expanding at $V_{\rm exp}=13\pm3$~\kms\
(its northern side approaching us and the southern one receding with
respect to the core).  Assuming a constant expansion velocity, its
kinematical age is then $4300\,D$~yrs, where $D$ is the distance in
kpc.  The systemic velocity of \object{A~79} is $-44\pm8$~\kms\ with
respect to the Local Standard of Rest.  No detailed information is
available for the expansion pattern in the faint, asymmetrical lobes.

In \object{He~2-428}, the long slit was oriented very close to the
minor and major axes of the ring.  As expected for an inclined
expanding ring, no velocity difference was found between regions of
the ring on either side of the central star along the major axis
($\mathrm{PA}=80\degr$), while a difference of $\Delta V=28$~\kms\ was
found along the minor axis ($\mathrm{PA}=-10\degr$).  Correcting for
an inclination $i=68\degr$ ($V_{\rm exp}=\Delta V/2\,\sin i$), the
deprojected expansion velocity of the ring is found to be 15~\kms,
with the northern side receding and the southern one approaching us
with respect to the core.  Its kinematical age is $2000\,D$~yrs, where
$D$ is the distance in kpc.  The systemic velocity of
\object{He~2-428}, as computed from the symmetry of the velocity field
of the ring, is $70\pm8$~\kms\ in the frame of the Local Standard of
Rest.  The lobes show hints of expansion velocities larger by
$\sim20$~\kms\ than those in the ring, as usually found in bipolar
PNe, but the signal is faint and no detailed spatiokinematical
modeling was possible.

In the core of \object{He~2-428}, H$\alpha$ is found to have a
relatively broad, double-peaked and asymmetrical profile.
The full-width at zero intensity of the lines is 137~km~s$^{-1}$.  
This profile might reflect a complex velocity field in the unresolved
circumstellar nebula, but its double peak, if not the asymmetry,
could be the consequence of transfer effects in a static stellar envelope
(Magnan \& de Laverny \cite{mag97}).

The expansion velocities of the rings of \object{He~2-428} and
\object{A~79} are low when compared with the usual expansion
velocities of bipolar PNe, which go up to several hundred \kms\
(Corradi \& Schwarz \cite{CS95}).  However, these high velocities are
found in the polar directions of bipolar PNe, whereas the equatorial
velocities of these objects are generally much lower (e.g. Corradi \&
Schwarz \cite{CS93}), and similar to the velocities found for
\object{A~79} and \object{He~2-428}.

\section{Discussion}
\label{disc}

The distances to \object{A~79} and \object{He~2-428} have been derived
assuming that their measured systemic velocities reflect their
participation in the general circular rotation around the Galactic
centre.  This is expected to be a good assumption for most bipolar
PNe, since many of the characteristics of these objects relate them to
the Galactic young disc population (Corradi \& Schwarz \cite{CS95}),
but it could fail for
\object{He~2-428} if -- as its low abundances may imply -- this nebula is
related to an older population.

For a standard Galactic rotation curve (cf. Corradi \& Schwarz
\cite{CS95}), the distance to \object{A~79} is found to be
5$\pm$2~kpc, with the error taking into account the uncertainty in its
derived systemic velocity and the dispersion in the velocity
ellipsoids at a given location in the Galaxy.  An independent estimate
of the distance to \object{A~79} can be obtained assuming that its
central star belongs to the nebula.  We have performed synthetic
photometry on our low-resolution spectrum of this star to derive its
$V$ magnitude.  Correcting for a light loss of 50\% due to the narrow
slit width, we derive $V=16.7$~mag.  Assuming that the star is not a
supergiant and correcting the spectrum for a reddening $A_{\rm
V}=1.3\mbox{--}1.9$~mag (each value obtained from the comparison of
the observed spectrum with the spectra given by Jacoby et al.\
(\cite{jac84}) for stars of type F0~V, F0~IV, F0~III, F3~V or A9~V) we
obtain $V=14.8\mbox{--}15.4$.  These values, when compared with the
absolute magnitudes expected for the corresponding spectral types
(Schmidt-Kaler \cite{Sch82}) give a distance range 2.7--4.5~kpc, which
is consistent with our distance estimate for \object{A~79} given above
and thus provides marginal evidence for the association of the star
with the nebula.  With a distance of 3--5~kpc, the age of the nebular
ring of \object{A~79} would be 13\,000--21\,500~yrs, its radius
0.17--0.28~pc, and the total size of the nebula 1.2--2.0~pc.  Previous
distance determinations for this object give 1.8--4.5~kpc (Zhang
\cite{zh95} and references therein).

Assuming that the systemic velocity of \object{He~2-428} is also due
to its circular rotation around the Galactic centre, we find a
distance between 4 and 8 kpc.  For this distance range, the
kinematical age of the ring would be 8\,000--16\,000~yrs, its radius
0.12--0.24~pc and the total size of the nebula, including what is
observed of the faint lobes, 1--2~pc.  The possibility that
\object{He~2-428} is a type~III PN (see Sect.~\ref{abund}), a class
with peculiar velocities of $\sim64$~\kms\ (Costa et al.\
\cite{cos96}), casts some doubts on the derived parameters.  However,
previous distance determinations for this object, based on independent
methods, give a distance of 7.1--10.0~kpc (Zhang \cite{zh95} and
references therein); the lower limit is consistent with our derived
distance range and the upper limit implies negligible changes in the
age and size derived for the nebula.

The ages derived for \object{A~79} and \object{He~2-428}, $\ge10^4$~yrs,
imply that they are evolved PNe.
At such ages, the stellar remnants are expected to have evolved to high
temperatures and then to low luminosities in their post-AGB evolution, unless
they have low masses.
The large \element{He} and \element{N} abundances of \object{A~79}
suggest a high-mass progenitor (cf. Corradi \& Schwarz
\cite{CS95}),
but assuming that the star at its centre belongs to the system, its
relative brightness and spectral type (F0) do not match those of an
aged post-AGB nucleus.  In the case of \object{He~2-428}, the central
star is hidden within a compact gaseous envelope, partly neutral with
very high densities ($N_{\rm e}>10^{10}$~cm$^{-3}$ in the \ion{Ca}{ii}
emitting region) and partly ionized with lower but still notable
densities ($N_{\rm e}$ from $10^{3.3}$ to $10^{6}$~cm$^{-3}$, see
Sect.~\ref{nhe2428}).  The presence of this dense circumstellar
envelope is clearly inconsistent with a post-AGB star 10$^4$~yrs after
the ejection of the PN.

It has been proposed that the nuclei of proto-PNe and young PNe with
anomalous spectral types, unable to produce the ionization observed in
their nebulae, are rapidly evolving single stars that experience rapid
increases of their temperature (up to a level at which the surrounding
nebula gets ionized) following sporadic post-AGB mass loss episodes
(see e.g. Bobrowsky et al.\ \cite{bob98}).  However, this cannot be
the case for \object{A~79} or \object{He~2-428} because this
scenario is expected to apply to the early phases of the post-AGB
evolution, whereas these objects have quite old nebulae.

The simplest explanation of these observations is then that the
central stars of both \object{He~2-428} and \object{A~79} are
binaries.  In the former, the nebular core would be produced by mass
transfer effects in an interacting binary system, leading to the
formation of a dense circumbinary nebula and possibly an accretion
disc (e.g. Morris \cite{mor87}, Soker \& Rappaport \cite{sok01}).
In fact, \object{He~2-428} might be similar to the
symbiotic binaries \object{R~Aqr} and \object{V417~Cen} (Solf \& Ulrich
\cite{So85}, van Winckel et al. \cite{van94}), which in addition of having a
dense circumbinary core do also have bipolar/ring nebulae as \object{He~2-428}
(the similarity with the nebula of \object{V417~Cen} is especially striking).
This would imply the presence in \object{He~2-428} of a cool star with a
strong wind, which could explain the red excess found in the spectrum
(see Sect.~\ref{nhe2428}).

As for \object{A~79}, the absence of nebular emission from the core suggests
that no strong winds and mass transfer/exchange between the two
stellar components are now present (this is expected for a system
composed of an F star plus an aged post-AGB companion providing the
ionization of the nebula).  A similar case might be \object{NGC~2346}, which
is known to have a binary central star (Feibelman \& Aller \cite{fei83}).
Note that the presence of a binary system would also offer a
natural explanation for the collimation of the large bipolar nebulae
of \object{He~2-428} and \object{A~79} (Soker \& Rappaport \cite{sok01}).

We have no constraints on the age of the nebula in \object{M~1-91} and
we cannot therefore exclude that it is very young and fits the
single-star scenario above. However, the properties of its sibling
\object{M~2-9} also point to the presence of a binary system to produce the
variety of its outflows (Doyle et al. \cite{doy00}; Livio \& Soker
\cite{liv01}). The similarity between \object{M~1-91} and
\object{M~2-9} is not limited to the overall morphology.
The available distances for \object{M~1-91}, of 7.0 or 7.8~kpc (Acker
et al.\ \cite{ack92}), imply a height over the Galactic plane
$z=440\mbox{--}490$~pc, to be compared with $z=309$~pc for
\object{M~2-9} and a mean $\langle|z|\rangle=130$~pc for a sample of
35 bipolar PNe (Corradi \& Schwarz \cite{CS95}).  The chemical
compositions of \object{M~1-91} and \object{M~2-9} are also similar,
both at the lower limit or below the abundance range spanned by other
bipolar PNe, as can be seen in Table~8, where the abundances shown for
\object{M~2-9} have been derived from the line intensities given
by Barker (\cite{bar78}) and the same atomic data and ionization
correction factors used for the other objects (see also Corradi \&
Schwarz 1995).

Finally, note that the departure from axisymmetry observed in the
lobes of \object{He~2-428} and \object{A~79} (see Fig.~\ref{fimag}),
as well as in the symbiotic nebula \object{V417~Cen}
(van Winckel et al.\ \cite{van94}), are all very similar to each other.
This suggests that the asymmetry is not just due to irregular
mass loss (e.g. caused by local mass loss events on the progenitor
surface). A binary system (especially with an eccentric orbit) would
also offer an explanation for the formation of such asymmetries (Soker
\& Rappaport \cite{sok01}).

\section{Summary and conclusions}

We have presented a spectroscopical study of the bipolar PNe \object{A~79},
\object{He~2-428} and \object{M~1-91}, with the following results:
\begin{itemize}
\item
\object{He~2-428} and \object{M~1-91} have high density (from $10^{3.3}$ to
$10^{6.5}$~cm$^{-3}$) unresolved nebular cores that indicate that
strong mass loss/exchange phenomena are occurring close to their
central stars;
\item
an additional region with densities larger than $10^{10}$~cm$^{-3}$ is
revealed in the core of \object{He~2-428} by the presence of the \ion{Ca}{ii}
infrared triplet in emission, possibly originating in an (accretion?)
disk;
\item
an F0 star is found at the symmetry centre of the nebula of
\object{A~79}.  Its reddening and distance are consistent with the
star being physically associated with the system, but radial velocity
measurements of the stellar features would be needed to confirm this
association;
\item
the continuum emission and the absorption lines in the core of \object{He~2-428}
can be reproduced with a hot star and an additional continuum component,
which is probably due to a late-type stellar companion;
\item
while \object{A~79} is overabundant in \element{N} and \element{He},
\object{He~2-428} and \object{M~1-91} have $\element{N}/\element{O}$
abundance ratios one order of magnitude lower, but still within the range
spanned by bipolar PNe; 
\item
\object{He~2-428} shows rather low abundances of all elements relative to
\element{H} (excluding \element{He}), which suggests that it
has arisen from a low-mass progenitor. This conclusion is strengthened
by the absence of \element{H}$_2$ emission in the nebula
(Guerrero et al.\ \cite{gue00});
\item
the nebulae of \object{A~79} and \object{He~2-428} are composed of
tilted equatorial rings expanding at a velocity of $\sim$15~\kms, from
which faint and irregular lobes depart. The nebulae are evolved, with
ages $\ge$10$^4$~yrs.
\end{itemize}

The associations of these old nebulae with a dense nebular core
(\object{He~2-428}) or with a central star with F0 spectral type
(\object{A~79}) are interpreted as evidence for the binarity of their
nuclei. In the literature, another dozen of bipolar PNe are known to
possess very dense nebular cores or central stars with anomalous
spectral types (see Corradi
\cite{cor95}). They provide prime targets for future observational
studies, by means of radial velocity monitoring or IR and UV imaging
and photometric searches for companions, aimed at testing the idea
that {\em all} bipolar PNe contain binary systems.

\begin{acknowledgements}
We thank M.~Guerrero for taking the INT spectra of He~2-428 and M~1-91
during service time.  We also thank D.~Mayya for revising the English text and
D.~Mayya, J.~Mikolajewska and D.~Pollacco for useful comments and discussions.
We thank the referee, R.~Tylenda, for his suggestions, which have helped to
improve this paper. This research has
made use of NASA's Astrophysics Data System Bibliographic Services and
the SIMBAD database, operated at CDS, Strasbourg, France.  A grant of
the Spanish DGES PB97--1435--C02--01 provided partial support for this
work.
\end{acknowledgements}


\end{document}